\documentclass[10pt,twocolumn,prd,aps,superscriptaddress,floatfix,nobalancelastpage,
preprintnumbers,reprint]{revtex4-2}

\usepackage[dvipsnames]{xcolor}
\usepackage[colorlinks=true,breaklinks=true]{hyperref}
\hypersetup{allcolors=[rgb]{0.0 0.0 0.70},linkcolor=blue}
\usepackage{orcidlink}
\usepackage{microtype}

\usepackage{amsmath}
\usepackage{amsfonts}
\usepackage{amssymb}
\usepackage{mathtools}
\usepackage{multirow}
\usepackage{bm}
\usepackage[utf8]{inputenc}
\usepackage{graphicx}
\usepackage{dcolumn}
\usepackage{bm}
\usepackage{comment}
\usepackage{multirow}
\usepackage{booktabs}
\usepackage[normalem]{ulem}

\usepackage{color}
\begin{document}

\title{Neutrino Flavor Conversion Shapes the Rate of Failed Core-collapse Supernovae}

\author{Mariam Gogilashvili \orcidlink{0000-0002-6944-8052}}
 \email{mariam.gogilashvili@nbi.ku.dk}
 
\author{Irene Tamborra \orcidlink{0000-0001-7449-104X}}
 \email{tamborra@nbi.ku.dk}
\affiliation{Niels Bohr International Academy and DARK, Niels Bohr Institute, University of Copenhagen, Blegdamsvej 17, 2100, Copenhagen, Denmark
}

\begin{abstract}
The relative rate of neutron stars and  black holes produced by  the collapse of massive stars is highly uncertain.
We simulate the stellar collapse of $195$ progenitors with masses between $9\, M_\odot$ and $120\, M_\odot$, incorporating a schematic treatment of neutrino flavor conversion. 
We find that flavor transformation    reshapes the  explodability of massive stars---especially in the $16$--$30\, M_\odot$ mass range---and modifies the compact remnant mass distribution. Our findings identify neutrino flavor conversion as a fundamental ingredient in predicting  neutron star and black hole populations, while naturally easing  the 
red-supergiant and the supernova-rate problems, as well as reconciling theoretical expectations with  the low-mass tail of the observed neutron star mass distribution. 
\end{abstract}

\maketitle

%\tableofcontents
%%%%%%%%%%%%%%%%%%%%%%%%%%%%%%%%%%%%%%%%%%%%%%%%%%%%%%%%%%%%%%%%%%%%%%%%%%%%%%%
% Introduction
%%%%%%%%%%%%%%%%%%%%%%%%%%%%%%%%%%%%%%%%%%%%%%%%%%%%%%%%%%%%%%%%%%%%%%%%%%%%%%%

\textbf{\textit{Introduction.}}---
When a  star with mass $M \gtrsim 8\,M_\odot$ 
exhausts its nuclear fuel, its iron core undergoes gravitational collapse.  If neutrino heating successfully revives the stalled shock wave, the star explodes as a core-collapse supernova (CCSN), leading to the formation of  
a neutron star~\cite{Bethe:1985sox,BURROWS1986}. Otherwise, if the star fails to explode, it 
collapses, forming a black hole~\cite{FISCHER2009,OCONNOR2011,Burrows:2024wqv}. Determining which massive stars explode as CCSNe and which collapse  into black holes remains a major unsolved problem in astrophysics, 
with direct implications for  the birth rates and mass 
distributions of neutron stars  and black holes,
 nucleosynthesis, and stellar feedback. 

It was initially believed that black hole formation could occur only for progenitor stars with mass exceeding $40\, M_\odot$~\cite{Fryer:1999mi}, affecting  the distribution of neutron star and black hole masses~\cite{Belczynski:2011bn,Fryer:2011cx,Fryer:1999ht,Fryer:1999mi}.  
However, recent work, based on  spherically symmetric simulations of the stellar collapse, has jeopardized such understanding,  revealing that the CCSN explodability is  non-monotonic with the progenitor mass, and black holes can originate from CCSN progenitors with mass as low as $13\,M_\odot$~\cite{Ugliano2012, Ertl2016, Sukhbold2016, Burrows2019, Boccioli2024_remnant}.

On the observational side, growing evidence suggests that massive stars may 
undergo failed explosions, leading to optically dark or faint 
transients~\cite{Kochanek:2008mp, Kochanek:2013yca,Kochanek:2014mwa,De:2024qqx,Kochanek:2023aob,Adams:2016hit,Fores-Toribio:2026wvr}. These findings  suggest that 
about $5$--$50\%$ of all collapses fail to produce a visible explosion, consistent 
with  theoretical expectations~\cite{Ugliano2012, Horiuchi:2014ska, Ertl2016,Sukhbold2016, Burrows2019,Boccioli2024_remnant} and the preliminary exclusion limits on the fraction of failed explosions  from the diffuse supernova neutrino background~\cite{Super-Kamiokande:2021jaq,Super-Kamiokande:2025sxh,Lien:2010yb,Martinez-Mirave:2024zck}.
Failed CCSNe may contribute to address the red supergiant 
problem (i.e., the absence of red supergiant progenitors in the mass range 
$16$--$30\,M_\odot$ among identified Type IIP supernova 
progenitors)~\cite{Smartt2009,Davies2020} as well as the CCSN rate 
problem (i.e., the fact that the observed cosmic CCSN rate falls systematically 
below the rate inferred from the cosmic star formation 
rate)~\cite{Horiuchi:2011zz}. 
In addition, radio pulsar and X-ray binary observations provide evidence for a 
neutron star mass distribution in the range $1.2$--$2.1\,M_\odot$, with 
a  dearth of black 
holes below $5\,M_\odot$~\cite{Farr:2010tu,Ozel:2016oaf,
Antoniadis:2016hxz, Alsing:2017bbc,You:2024bmk}. These constraints have recently been complemented by the detection of gravitational waves from 
compact binary mergers~\cite{Abbott2023_GWTC3,LIGOScientific:2025slb,LIGOScientific:2026wxz}. Accounting for these observational constraints  
simultaneously  challenges    theoretical CCSN
models~\cite{Fryer:2011cx,Sukhbold2016,Boccioli2024_remnant, Muller:2018utr}, suggesting 
that key physics, governing the explosion mechanism and determining which progenitors fail to explode, is yet to  be established. 

The CCSN explosion mechanism is driven by  neutrinos. The latter are weakly interacting elementary particles that carry away about $99\%$ of the gravitational binding energy [$\mathcal{O}(10^{53})$~erg] of the proto-neutron star formed during the core collapse. Electron neutrinos and antineutrinos mediate the energy transfer behind 
the stalled shock through charged-current interactions~\cite{Bethe:1985sox};  
whether their energy deposition is sufficient to revive the shock ultimately 
determines the fate of the collapse.

Despite its crucial role  in driving the fate of the explosion, neutrino physics in CCSNe is not fully understood. In particular, state-of-the-art multi-dimensional hydrodynamic simulations of CCSNe do not account for the fact that neutrinos can change their flavor while propagating in the CCSN core (see, e.g., Refs.~\cite{Mezzacappa2020,Duan2010,Mirizzi2016,Tamborra2021,Tamborra2025,Volpe2024,Johns2025} for an overview). The assumption that neutrino flavor conversion (FC) could be neglected in the modeling of the explosion mechanism  was justified by the understanding that  FC would have a negligible effect behind the shock~\cite{Dasgupta2012}. However, recent insights into the physics of neutrino self-interaction question this picture as they show that  FC  can take place even in the surroundings of the proto-neutron star~\cite{Sawyer2005,Sawyer2016,Chakraborty2016,Izaguirre2017,Shalgar2023,Johns2023,Fiorillo2025}. If neutrino FC occurs in the  CCSN core, it can affect the neutrino-driven delayed explosion mechanism, as suggested by preliminary work accounting for  FC in CCSN simulations using  schematic approaches~\cite{Ehring2023, Ehring2023abs,Nagakura2023,Wang2025,Akaho2026,Mori2025, Gogilashvili2026inprep}.

In this  paper, for the first time, we explore whether FC affects  the  rate of failed CCSNe and the properties of the compact remnants. For this purpose, we build on  Ref.~\cite{Gogilashvili2026inprep}, that employs the  instantaneous flavor equilibration of (anti)neutrinos from Ref.~\cite{Ehring2023}  in 1D+ CCSN  simulations performed with \texttt{GR1D}~\cite{OConnor2010, OConnor2015, Boccioli2021_STIR_GR}. We investigate the impact of FC on the fate of the CCSN explosion using a set of $195$  progenitors with solar metallicity from Ref.~\cite{Sukhbold2016} with masses between $9\, M_\odot$ and $120\, M_\odot$. 

Our results suggest that  FC  increases the 
fraction of failed explosions, especially in the  $16$--$30\,M_\odot$ mass range, potentially helping to address the supernova rate problem. Moreover, when FC occurs near the stalled shock, it accelerates shock 
revival, yielding less massive neutron stars, in better agreement with the observed mass distribution.

%%%%%%%%%%%%%%%%%%%%%%%%%%%%%%%%%%%%%%%%%%%%%%%%%%%%%%%%%%%%%%%%%%%%%%%%%%%%%%%
% Model
%%%%%%%%%%%%%%%%%%%%%%%%%%%%%%%%%%%%%%%%%%%%%%%%%%%%%%%%%%%%%%%%%%%%%%%%%%%%%%%
\textbf{\textit{Model Setup.}}---We model the  stellar collapse using the open-source, spherically symmetric code \texttt{GR1D}~\cite{OConnor2010, OConnor2015}, which evolves general-relativistic hydrodynamics coupled to energy-dependent neutrino transport. As detailed in Ref.~\cite{Gogilashvili2026inprep}, our simulations employ a radial grid with $850$ zones, resolving the innermost $20\, \mathrm{km}$ with uniform spacing and increasing logarithmically at larger radii. The hydrodynamics is solved using a finite-volume approach with second-order Runge–Kutta time integration, capturing both the collapse of the iron core and the post-bounce evolution. Neutrino transport is treated with an energy-dependent moment formalism using M1 closure for three species ($\nu_e$, $\bar{\nu}_e$, and $\nu_x = \bar\nu_x$, with $x= \mu$ or $\tau$) over $18$ energy bins. Neutrino–matter interactions are modeled using tabulated rates from \texttt{NuLib}~\cite{OConnor2015}.
To account for the effects of neutrino-driven convection and turbulence, we employ
the Supernova Turbulence in Reduced-dimensionality (\texttt{STIR}) model~\cite{Couch2020_STIR, Boccioli2021_STIR_GR}, with
the turbulence strength set by the dimensionless parameter $\alpha_{\rm MLT} = 1.51$.

Flavor conversion is taken into account using the parametric  scheme proposed in Refs.~\cite{Ehring2023, Just2022}, which instantaneously leads to flavor equipartition, while conserving electron lepton number as well as the number and momentum of (anti)neutrinos for each energy bin. We apply the equipartition scheme  to each energy bin shortly after bounce ($t = 0.02\,\mathrm{s}$) in regions where the matter density is below a threshold baryon density $\rho_c$, varied between $10^{13}$ and $10^9\,\mathrm{g/cm^{3}}$, i.e., spanning the radial region between the neutrinosphere and the stalled shock. The critical density $\rho_c$ serves as a proxy for the uncertain radial
extent of flavor conversion (cf., e.g., Fig.~2 of Ref.~\cite{Tamborra2025}).
As illustrated in Fig.~1 of
Ref.~\cite{Gogilashvili2026inprep}, the four thresholds correspond to progressively
deeper regions of the post-bounce core: $\rho_c=10^{9}\,\mathrm{g/cm^3}$
lies in the gain region near the shock, $\rho_c=10^{10}\,\mathrm{g/cm^3}$
close to the gain radius, $\rho_c=10^{11}\,\mathrm{g/cm^3}$ near the
neutrinosphere, and $\rho_c=10^{13}\,\mathrm{g/cm^3}$ deep within the
proto-neutron star. The exact radius corresponding to each critical density evolves in
time and depends on progenitor structure and equation of state, but its qualitative ordering holds across
our progenitor set. 
By modifying the relative spectra of electron neutrinos, electron antineutrinos, and heavy-lepton neutrinos, FC  affects charged-current heating and cooling in the  region behind the shock, thereby altering the conditions for shock revival~\cite{Gogilashvili2026inprep,Ehring2023,Ehring2023abs}. We refer the interested reader to Ref.~\cite{Gogilashvili2026inprep} for additional  details.

We simulate $195$ progenitors from Ref.~\cite{Sukhbold2016}, spanning zero-age main-sequence (ZAMS) masses from $9\, M_\odot$ through   $120\, M_\odot$. This  selection allows us to capture the diversity of pre-collapse cores, from low-mass stars with steep density gradients to more massive stars with extended, loosely bound envelopes.
All CCSN models are simulated using the Steiner, Fischer, and Hempel  (SFHo) equation of state~\cite{STEINER2013}.  Each progenitor is evolved from the onset of collapse through $1$--$3\, \mathrm{s}$ after core bounce. 

\begin{figure*}
    \centering
    \includegraphics[width=\textwidth]{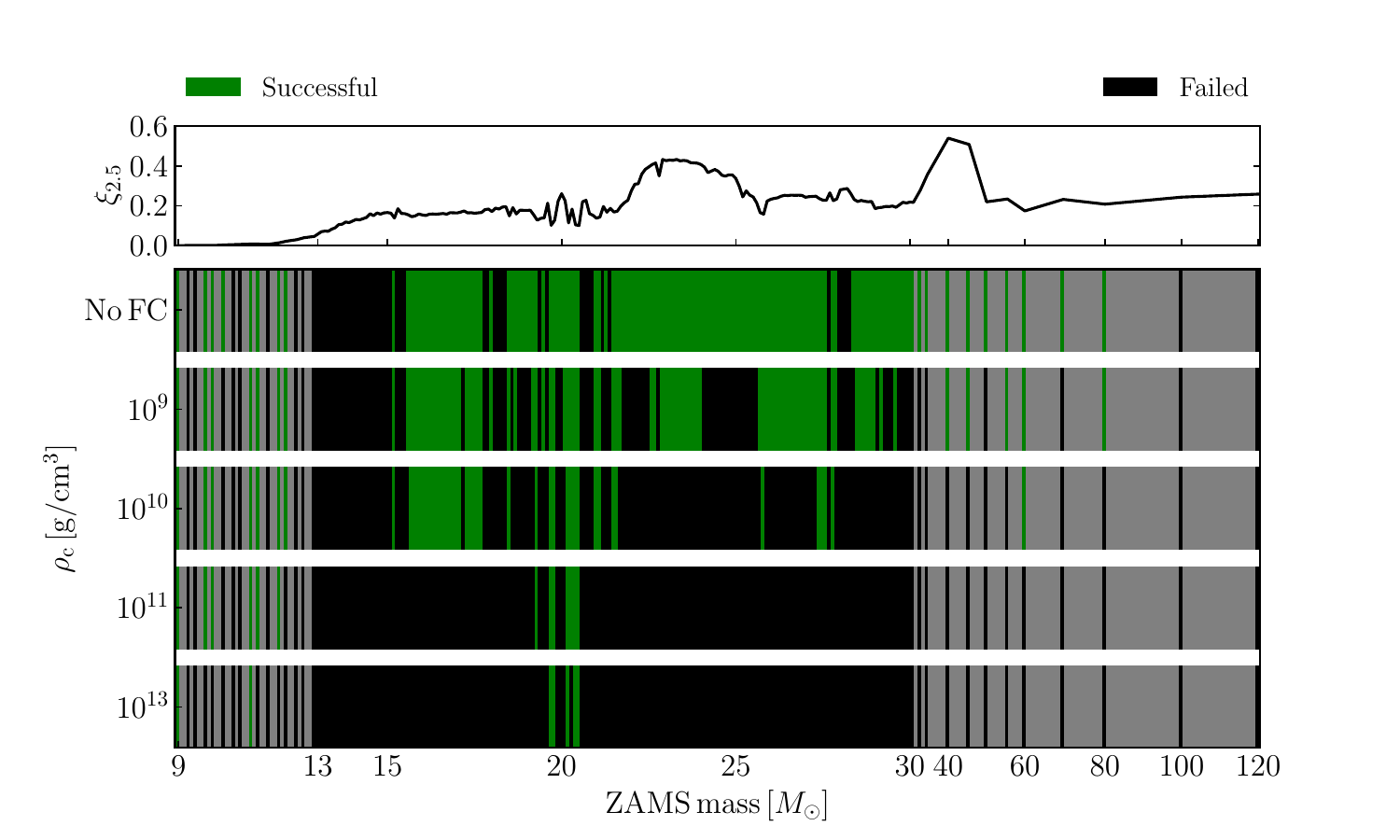}
    \caption{Islands of explodability for CCSNe. {\it Top panel:} Pre-supernova core compactness ($\xi_{2.5}$) as a function of the ZAMS mass showing the correlation between progenitor structure and CCSN explodability. {\it Bottom panel: } Islands of explodability as  functions of the ZAMS mass. Each horizontal band displays successful (green) or failed (black) CCSN explosions;  the gray bars 
    denote progenitors that are not simulated. The top row represents the reference case without FC. From the second to the bottom row, respectively, FC is triggered  below a critical baryon density: $\rho_c = 10^{9}, 10^{10}, 10^{11}$, and $10^{13}$\,g/cm$^3$.   Failed explosions are favored when FC is taken into account, especially in the $16$--$30\, M_\odot$ mass range.}
\label{fig:islands_of_explodability}
\end{figure*}

To characterize the explodability potential of CCSNe (cf., e.g., Refs.~\cite{Ugliano2012,Ertl2016,Sukhbold2016,Maltsev:2025bgs} for dedicated work in this direction), we use  the progenitor compactness parameter~\cite{OCONNOR2011}: 
    $\xi_M = {(M/M_\odot)}/{[R(M)/1000\, \mathrm{km}]}$,
where $R(M)$ is the radial coordinate enclosing a mass $M$. The  compactness of  our models is evaluated at $M=2.5\, M_\odot$ just prior to collapse  and is displayed in the top panel of Fig.~\ref{fig:islands_of_explodability}.

%%%%%%%%%%%%%%%%%%%%%%%%%%%%%%%%%%%%%%%%%%%%%%%%%%%%%%%%%%%%%%%%%%%%%%%%%%%%%%%
% Results
%%%%%%%%%%%%%%%%%%%%%%%%%%%%%%%%%%%%%%%%%%%%%%%%%%%%%%%%%%%%%%%%%%%%%%%%%%%%%%%
%\section{Results}\label{sec:Results} 
\textbf{\textit{Fraction of Failed Supernovae.}}---Figure~\ref{fig:islands_of_explodability} presents our results on the explodability landscape 
for  $195$ progenitors with masses between $9\,M_\odot$ and $120\,M_\odot$, simulated 
 for five scenarios: without FC (No~FC) and with FC  imposed below the critical densities $\rho_\mathrm{c} = 10^{9}$, 
$10^{10}$, $10^{11}$, and $10^{13}\,\mathrm{g/cm^{3}}$, from top to bottom, respectively. 
We assume that core collapse leads to a successful explosion (green bars) if the shock radius exceeds $1000 \, \mathrm{km}$, otherwise  a failed supernova occurs (black bars). 

\begin{table}[t]
\centering
\caption{Fraction of the failed explosions for models without and with FC instantaneously  triggered below the baryon density $\rho_c$. We list the  
 fraction of our models that fails to explode ($f_{\rm fail}$) as well as the fraction of failed explosions weighted by the Salpeter  IMF ($f_{\rm Salp}$).}
\label{tab:collapse_fractions}
\begin{ruledtabular}
\setlength{\tabcolsep}{12pt}
\begin{tabular}{lcc}
$\rho_c$ [g/cm$^3$] & $f_{\rm fail}$ & $f_{\rm Salp}$\\
\hline
No FC    & 25.6\% & 27.0\% \\
$10^{9}$  & 50.8\% & 47.6\% \\
$10^{10}$ & 77.4\% & 61.9\% \\
$10^{11}$ & 93.3\% & 73.9\% \\
$10^{13}$ & 96.4\% & 88.3\% \\
\end{tabular}
\end{ruledtabular}
\end{table}
Overall, we  see that FC increases the fraction of failed explosions, especially in the  $16$--$30\,M_\odot$ mass range.
Figure~\ref{fig:islands_of_explodability} also suggest that the fraction of failed collapses increases with increasing $\rho_\mathrm{c}$. 
Hence, 
the deeper inside the proto-neutron star flavor 
equilibration is achieved, the greater the suppression of shock revival. 
This trend  reflects the overall tendency of  FC in  
redistributing energy from electron-flavor neutrinos  to heavy-flavor neutrinos,  
with a consequent decrease of the efficiency of neutrino energy deposition~\cite{Gogilashvili2026inprep}. 

Without FC, $25.6\%$ of our CCSN models fail to explode, in agreement  with the existing literature based on spherically symmetric CCSN simulations~\cite{OCONNOR2011,Sukhbold2016, Boccioli2024_remnant,Ertl2016,Ugliano2012}. 
If we account for the distribution of CCSN progenitors according to the Salpeter  initial mass function (IMF)~\cite{Salpeter:1955it}, 
the fraction of failed  explosions
is $27\%$ in the absence of FC. This fraction is consistent with observational 
constraints~\cite{Kochanek:2008mp, Kochanek:2013yca,Kochanek:2014mwa,De:2024qqx,Kochanek:2023aob,Adams:2016hit,Fores-Toribio:2026wvr}.
When FC is taken into account,  the IMF-weighted fraction of failed explosions increases, ranging from $47.6\%$ for $\rho_c = 10^9\, \mathrm{g}/\mathrm{cm}^3$ to $88.3\%$ for $\rho_c = 10^{13}\, \mathrm{g}/\mathrm{cm}^3$, 
as summarized in Table~\ref{tab:collapse_fractions}. Note that the IMF proposed in Ref.~\cite{Kroupa:2001jy} leads to  similar results (not shown here). 

The changes induced by FC in the intermediate ZAMS mass range may contribute to alleviate the  
 red supergiant problem~\cite{Smartt2009, Davies2020} as well as the supernova rate 
problem~\cite{Horiuchi:2011zz}.  
However,  for  $\rho_\mathrm{c} = 10^{11}$ and  
$10^{13}\,\mathrm{g/cm^{3}}$, the IMF-weighted fractions of failed collapses are $73.9\%$ and $88.3\%$, seemingly in tension with observational constraints. These quantitative findings should be interpreted with caution. 
To compute the the exact fraction of failed collapses, FC should be  incorporated in CCSN simulations dynamically, accounting for   
the temporal evolution and spatial extent of FC.
To this purpose, advanced multi-dimensional subgrid models of FC---cf., e.g., Refs.~\cite{Xiong2021,Just2022,Padilla2022,Ehring2023,Zaizen2023,Zaizen2023a,Nagakura2024,Goimil2025,Liu2025,Xiong2025,Johns2025_subgrid} for recent efforts in this direction---should be embedded in CCSN simulations dynamically. Such subgrid models build on the understanding of FC coming from kinetic simulations of neutrino transport (currently still under development; see, e.g., Refs.~\cite{Nagakura2022,Shalgar2023,Shalgar2023a,Xiong2024,Cornelius2024,Shalgar2025,Grohs:2025ajr,Richers:2021nbx}). 
Moreover, our CCSN models do not account for proto-neutron star convection. However, FC triggered  at high $\rho_c$  is expected to boost convection in the hot proto-neutron star~\cite{Ehring:2024mjx}, with  a consequent  potential enhancement of neutrino heating. In addition, we model the CCSN population employing the SFHo nuclear equation of state, however, the impact of FC  on the fate of the explosion may be quantitatively affected if other nuclear equations of state were to be considered~\cite{Gogilashvili2026inprep,Akaho2026}.

In the absence of FC, our CCSN progenitors~\cite{Sukhbold2016}  
predict a large fraction of failed explosions in the range between  $13\,M_\odot$ and $16\,M_\odot$. Since low- and intermediate-mass progenitors are the most abundant
according to the IMF, they   drive the
overall rate of failed explosions. It should be noted, however, that the predicted explodability landscape is sensitive to  the choice of the progenitor set as well as the details of the physics implemented in the CCSN hydrodynamic simulation: different stellar evolution models (e.g.~Refs.~\cite{Woosley2002, Chieffi:2020gxh}) and different neutrino-driven explosion treatments (e.g.~Refs.~\cite{OCONNOR2011, Ertl2016, Muller:2018utr, Boccioli2023_explodability, Janka2025}) yield different distributions of failed and successful collapses, including in the $13$--$16\,M_\odot$ mass range. Furthermore, all progenitor models used in the literature on the topic are  one dimensional, assuming spherical symmetry in the pre-collapse stellar structure. Three-dimensional CCSN simulations suggest that multidimensional effects in the progenitor (such as convective shell burning  perturbations and asymmetries) can significantly alter the shock revival and thus the CCSN explodability~\cite{Bollig:2020phc,Whitehead2026}.

\begin{figure}
    \centering
\includegraphics[width=\columnwidth]{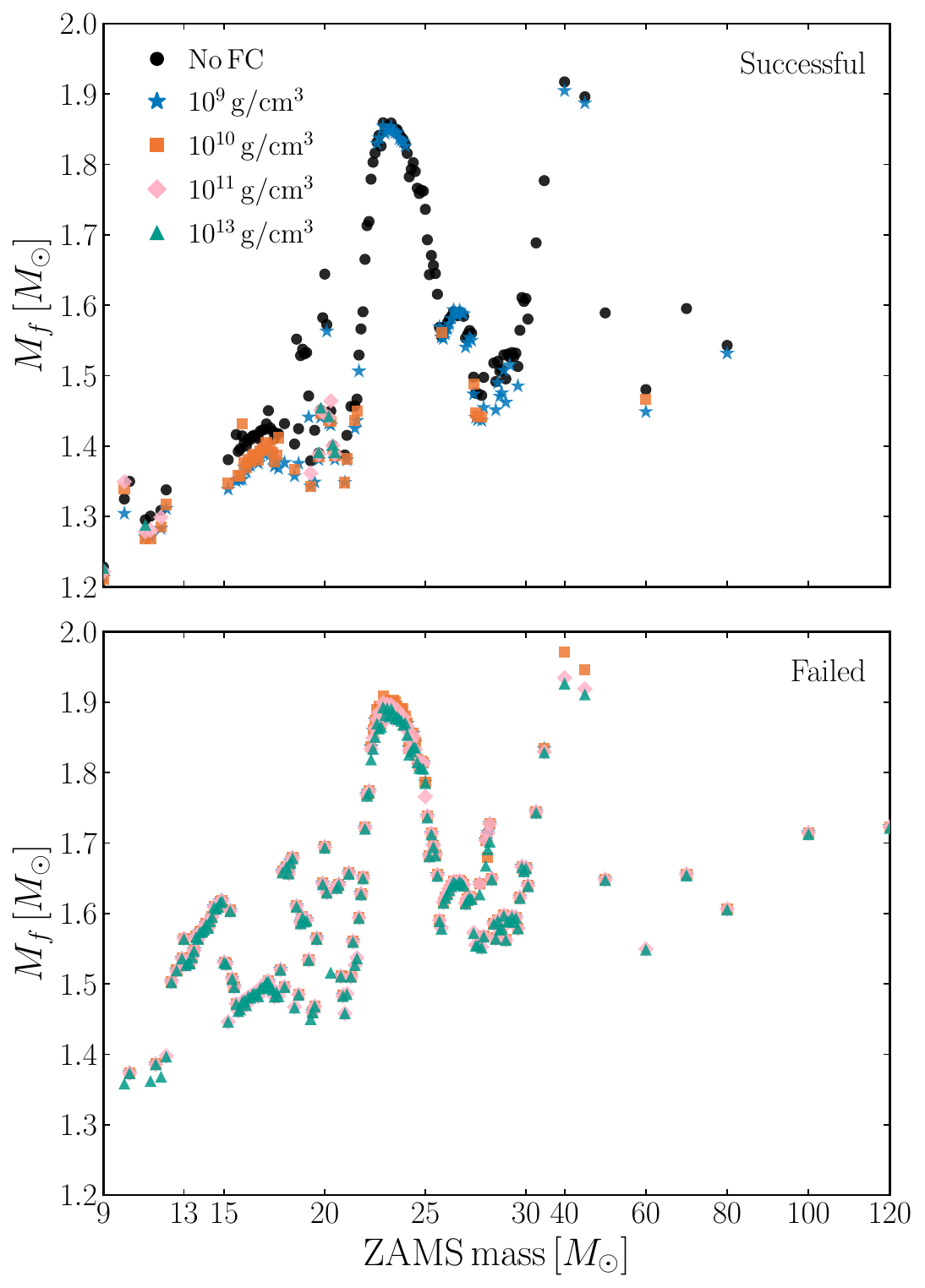}
    \caption{Baryonic mass of the compact remnant evaluated at  $t = 1$\,s post-bounce (or at the latest time step, if the black hole forms earlier) as a function of ZAMS mass. 
    The final mass of the remnant  is shown for the cases without FC (black circles) and with FC triggered below 
    $\rho_c = 10^{9}$ (blue stars), $10^{10}$ (orange squares), $10^{11}$ 
    (pink diamonds), and $10^{13}$\,g/cm$^3$ (teal triangles). The   progenitors that successfully explode (fail to explode) are represented in the upper (lower) panel. For 
    successful explosions, the compact remnant mass is systematically 
    smaller when FC is taken into account.
    Vice versa, when FC 
    suppresses or prevents shock revival, leading to black hole formation,  the remnant mass tends to be slightly  larger (black and blue symbols are underneath the orange ones).
    }
    \label{fig:NS_mass}
\end{figure}

\textbf{\textit{Compact Remnant Properties.}}---Flavor conversion also contributes to define the mass of the compact remnant, as shown in Fig.~\ref{fig:NS_mass}. The latter represents  the baryonic mass ($M_{f}$) enclosed within 
the energy- and flavor-averaged neutrinosphere radius, evaluated at 
$t = 1\, \mathrm{s}$ post-bounce or at the latest available time step for 
simulations that did not reach this time~\footnote{Our spherically symmetric simulations cannot capture the
simultaneous accretion-outflow phase seen in multi-dimensional
simulations, where continued accretion through downflows coexists
with outflows launched anisotropically. As a result, $M_f$ for
successful explosions should be interpreted as a lower bound on the
post-bounce mass growth.}.
For successful explosions (top panel of Fig.~\ref{fig:NS_mass}), 
FC  leads to  remnants more compact than those obtained  when FC is not taken into account (colored vs.~black points).
Triggering earlier and more energetic shock 
revival, FC is responsible for less efficient matter accretion onto the proto-neutron 
star, resulting in systematically lower  $M_{f}$. 
The impact of  FC on $M_{f}$ becomes more pronounced at lower 
$\rho_\mathrm{c}$, since the revival of the shock is 
 more efficient~\cite{Gogilashvili2026inprep}. 

For successful explosions,   $M_{f}$  spans 
the range between $1.2\, M_\odot$ and $1.9\, M_\odot$. Hence, after accounting for the 
baryonic to gravitational mass correction~\footnote{We convert baryonic to gravitational mass following $M_{\rm grav} = M_{\rm bary} - 0.075\,M_\odot\,(M_{\rm grav}/M_\odot)^2$~\cite{Timmes:1995kp}.},  FC is responsible for 
gravitational masses that agree better with the bulk of the observed 
neutron star population, peaking around $1.2$--$1.4\, M_\odot$~\cite{Ozel:2016oaf, Antoniadis:2016hxz,You:2024bmk}, compared to the systematically higher  neutron star masses
 obtained without FC. The lower gravitational mass obtained  with FC thus   eases the tension between CCSN theory and the observed low-mass end of the neutron star population~\cite{Muller:2024aod}.

For  failed explosions (bottom panel of Fig.~\ref{fig:NS_mass}), $M_{f}$  tends to be  
larger in the presence of FC,
as the absence of shock revival allows for
accretion onto the proto-neutron star to continue up to $t = 1 \,\mathrm{s}$ 
post-bounce (or until black hole formation, if the latter occurs earlier). 
This effect  is most pronounced for FC triggered at higher $\rho_c$, 
where shock revival is significantly delayed or  suppressed, 
sustaining accretion for longer. Nevertheless, the effective change in the baryonic mass of the compact remnant induced by FC seems to be less prominent than for the case of successful explosions. 
Note, however,  that $M_{f}$  should not be interpreted as the final black hole mass; the latter could be  substantially larger because of fallback accretion of the 
 stellar envelope, if   black hole formation occurs at $t \gtrsim 1\, \mathrm{s}$.

%%%%%%%%%%%%%%%%%%%%%%%%%%%%%%%%%%%%%%%%%%%%%%%%%%%%%%%%%%%%%%%%%%%%%%%%%%%%%%%
% Conclusions
%%%%%%%%%%%%%%%%%%%%%%%%%%%%%%%%%%%%%%%%%%%%%%%%%%%%%%%%%%%%%%%%%%%%%%%%%%%%%%%
\textbf{\textit{Conclusions.}}---We investigate, for the first time, the impact of 
FC on the explodability of $195$ CCSN progenitors with masses between 
 $9\,M_\odot$ and $120\,M_\odot$ using 1D+ hydrodynamic simulations. Our results show that 
FC has a profound  impact on the CCSN explodability. 
In fact, overall, FC tends to reduce the net energy deposition behind the stalled shock, 
systematically favoring failed explosions. 

We trigger FC  instantaneously below a critical baryon density ($\rho_{c}$), imposing  flavor equipartition while conserving the  electron lepton number as well as the number and momentum of neutrinos for each energy bin. As $\rho_\mathrm{c}$ increases, the fraction of failed explosions is larger due to more efficient  neutrino cooling~\cite{Gogilashvili2026inprep}.
Importantly,  
the models with mass in the  $16$--$30\,M_\odot$ range are particularly 
sensitive to FC, with a large number of otherwise successful explosions being 
converted into failures when FC is taken into account. 
The changes induced by FC in this mass range may help alleviate the 
 red supergiant problem~\cite{Smartt2009, Davies2020} as well as the supernova rate 
problem~\cite{Horiuchi:2011zz}.  
We stress that the relative increase of failed explosions when FC is taken into account is robust with respect to our reference CCSN progenitor population that does not  account for FC. However, our  fraction of failed collapses is expected to be  sensitive to the details of the simulation as well as  our schematic treatment of FC. Moreover,  the the \texttt{STIR} model of neutrino-driven
convection (based on the mixing-length approach),  captures the overall efficiency of
neutrino heating, but cannot reproduce nonradial dynamics such as
the standing accretion shock instability (SASI). SASI is expected to
be most relevant for compact, high-accretion-rate progenitors, and has
been found to aid shock expansion beyond what STIR alone reproduces at later post-bounce times~\cite{Couch2020_STIR}. The omission of SASI dynamics may bias our CCSN explodability landscape toward failed explosions, especially  for the most compact progenitors in our sample.

The  mass of the compact remnant left behind by successful explosions is also affected by FC. We find  systematically lower 
proto-neutron star masses in the CCSN models that account for FC,  reflecting earlier shock revival 
and reduced post-bounce accretion. 
Our findings imply that  FC should be taken into account in theoretical forecasts of the 
neutron star mass distribution, as gravitational-wave, X-ray  and pulsar-timing
observations establish a larger  sample of neutron star 
masses~\cite{Farr:2010tu,Ozel:2016oaf,
Antoniadis:2016hxz, Alsing:2017bbc, Kini:2026rjx,You:2024bmk}.

Neutrino FC is a key 
ingredient  in modeling the birth rates and mass distributions of 
 black holes and neutron stars. Advances in our understanding of the microphysics determining the collapse of massive stars are especially needed, as a growing number of multi-messenger observations challenges our understanding of the physics driving the latest stages of the life of massive stars.

%%%%%%%%%%%%%%%%%%%%%%%%%%%%%%%%%%%%%%%%%%%%%%%%%%%%%%%%%%%%%%%%%%%%%%%%%%%%%%%
%Acknowledgements %%%%%%%%%%%%%%%%%%%%%%%%%%%%%%%%%%%%%%%%%%%%%%%%%%%%%%%%%%%%%
%%%%%%%%%%%%%%%%%%%%%%%%%%%%%%%%%%%%%%%%%%%%%%%%%%%%%%%%%%%%%%%%%%%%%%%%%%%%%%%
\textbf{\textit{Acknowledgments.}}---
We thank Luca Boccioli, Hans-Thomas Janka and Evan P.~O'Connor  for insightful discussions.
This project has received support from the European Union (ERC, ANET, Project No.~101087058) and the Villum Foundation (Project No.~13164).
Views and opinions expressed are those of the authors only and do not necessarily reflect those of the European Union or the European Research Council. Neither the European Union nor the granting authority can be held responsible for them. We used the Tycho supercomputer hosted at the SCIENCE HPC center at the University of Copenhagen to perform the numerical simulations presented in this work.

\end{document}